# Securing Iris Templates using Combined User and Soft Biometric based Password Hardened Fuzzy Vault

V. S. Meenakshi

SNR Sons College,
Coimbatore,Tamil Nadu, India.
.

Dr. G. Padmavathi

Avinashilingam University for Women, Coimbatore,
Tamil Nadu, India.
.

*Abstract*—Personal identification and authentication is very crucial in the current scenario. Biometrics plays an important role in this area. Biometric based authentication has proved superior compared to traditional password based authentication. Anyhow biometrics is permanent feature of a person and cannot be reissued when compromised as passwords. To over come this problem, instead of storing the original biometric templates transformed templates can be stored. Whenever the transformation function is changed new revocable/cancelable templates are generated. Soft biometrics is ancillary information that can be combined with primary biometrics to identify a person in a better way. Iris has certain advantage compared to other biometric traits like fingerprint. Iris is an internal part that is less prone to damage. Moreover is very difficult for an attacker to capture an iris. The key advantage of iris biometrics is its stability or template longevity. Biometric systems are vulnerable to a variety of attacks. This work generates cancelable iris templates by applying user and soft biometric based password transformations and further secures the templates by biometric cryptographic construct fuzzy vault.

## I. INTRODUCTION

Biometrics is automated methods of identifying a person or verifying the identity of a person based on a physiological or behavioral characteristic. Biometric characteristics that have been commonly used are fingerprints, iris, face, hand, retina, voice, signature and keystroke dynamics. Biometrics identifies 'you as you'. Biometrics serves as an excellent alternative to traditional token or password based authentication methods. Biometric systems are excellent compared over traditional authentication methods. Biometric traits cannot be lost or forgotten and they are inherently more reliable. It is very difficult to copy, share and distribute a biometric trait. Biometric system requires the person being authenticated to be present at the time and point of authentication.

Anyhow biometrics cannot be revoked when they are spoofed. To overcome this cancelable biometric templates are generated that can be revoked when spoofed. Further they are secured by fuzzy vault construct.

### A. Merits of Iris

Iris is the colored ring surrounding the pupil of the eye. Iris biometric has certain merits compared to finger print. It is highly secure and uses a stable physiological trait. Iris is very difficult to spoof. Iris texture is different for right and left eye. They are unique even for identical twins. Iris is less prone to either intentional or unintentional modification when compared to fingerprint.

### B. Soft Biometrics

Soft biometrics provides ancillary information about a person (gender, ethnicity, age, height, weight, eye color etc). They lack distinctiveness or permanence. Hence Soft biometrics alone is not enough to differentiate two individuals. Anyhow when combined with primary biometrics (Fingerprint, Iris, Retina etc) soft biometrics gives better results.

### C. Cancelable Biometrics

Passwords can be revoked when it is stolen. Biometrics cannot be revoked when spoofed. This is the only disadvantage of biometrics as compared to passwords. Therefore instead of storing the biometrics as such transformed templates are stored. Whenever a transformed biometric template is spoofed another new template can be generated by changing the transformation function. This makes the biometric cancelable/ revocable similar to password. For different applications different transformation function can be used. This prevents the attacker to use the same captured template for other applications. Like passwords these transformed templates can be reissued on spoofing.

### D. Operation of Fuzzy Vault

Fuzzy vault is a cryptographic construct proposed by Juels and Sudan [2]. This construct is more suitable for applications where biometric authentication and cryptography are combined together. Fuzzy vault framework thus utilizes the advantages of both cryptography and biometrics. Fuzzy vault eliminates the key management problem as compared to other practical cryptosystems.

In fuzzy vault framework, the secret key S is locked by G, where G is an unordered set from the biometric sample. A polynomial P is constructed by encoding the secret S. This polynomial is evaluated by all the elements of the unordered set G.

A vault V is constructed by the union of unordered set G and chaff point set C which is not in G.
$$V = G \cup C$$
The union of the chaff point set hides the genuine point set from the attacker. Hiding the genuine point set secures the secret data S and user biometric template T.

The vault is unlocked with the query template T'. T' is represented by another unordered set U'. The user has to separate sufficient number of points from the vault V by comparing U' with V. By using error correction method the polynomial P can be successfully reconstructed if U' overlaps with U and secret S gets decoded. If there is not substantial overlapping between U and U' secret key S is not decoded. This construct is called fuzzy because the vault will get decoded even for very near values of U and U' and the secret key S can be retrieved. Therefore fuzzy vault construct become more suitable for biometric data which show inherent fuzziness hence the name fuzzy vault as proposed by Sudan [2].

The security of the fuzzy vault depends on the infeasibility of the polynomial reconstruction problem. The vault performance can be improved by adding more number of chaff points C to the vault.

### E.  Limitation of Fuzzy Vault Scheme

Fuzzy vault being a proven scheme has its own limitations [5].

(i) If the vault is compromised, the same biometric data cannot be used to construct a new vault. Fuzzy vault cannot be revoked. Fuzzy vault is prone to cross-matching of templates across various databases.

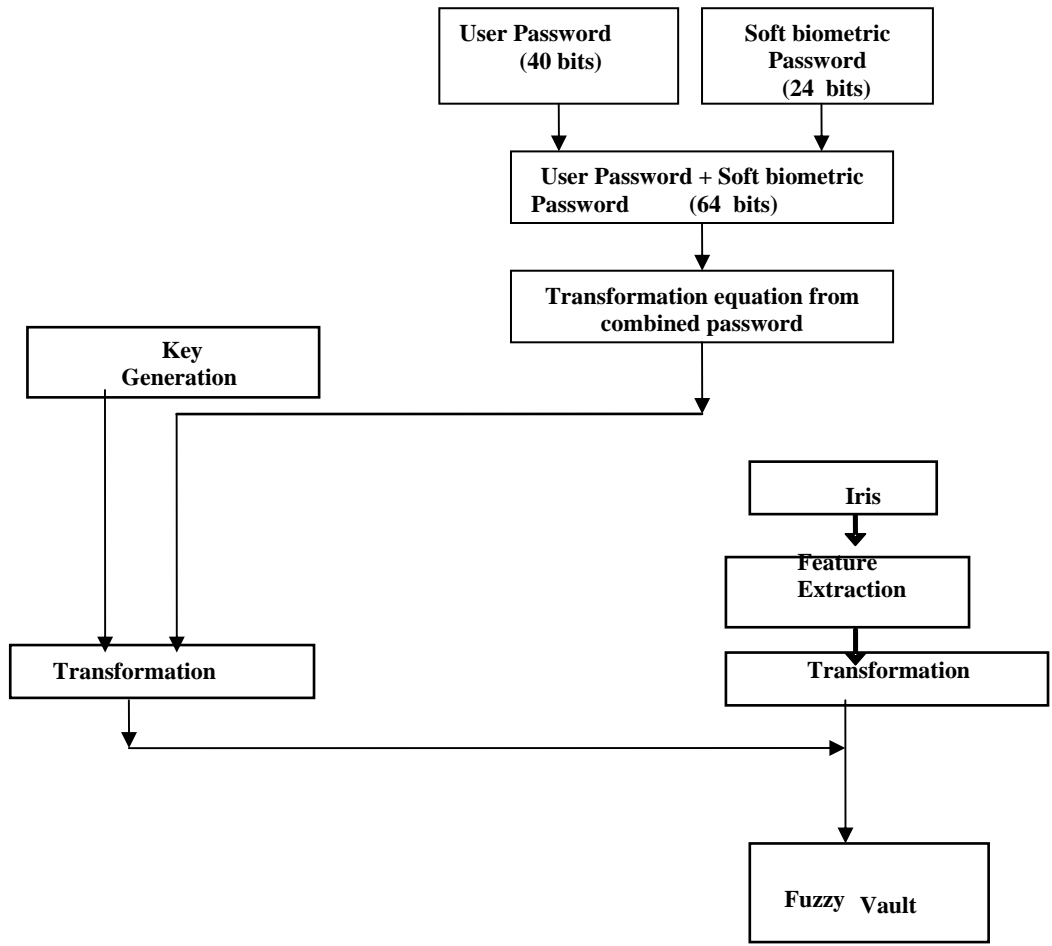

Fig. 1 Steps in combined password hardened fuzzy vault

(ii) Due to the non-uniform nature of the biometric features it is easy for an attacker to develop attacks based on statistical analysis of the points in the vault.

(iii) The vault contains more chaff points than the genuine points. This facilitates the attacker to substitute few points from his own biometric feature. Therefore the vault authenticates both the genuine user and the imposter using the same biometric identity. As a consequence, the false acceptance ratio of the system increases.

(iv) Original template of the genuine user is temporarily exposed. During this exposure the attacker can glean the template.

To overcome the limitations of fuzzy vault, password is used as an additional authentication factor. The proposed Iris fuzzy vault is hardened by combined user and biometric password. This enhances the user-privacy and adds an additional level of security.

*F. Steps in Combined Password Hardened Fuzzy Vault*

The hardened fuzzy vault overcomes the limitations of non-revocability and cross-matching by introducing an additional layer of security by password. If the password is compromised the basic security and privacy provided by the fuzzy vault is not affected. However, a compromised password makes the security level same as that of a fuzzy vault. Therefore, security of the password is crucial. It is very difficult for an attacker to compromise the biometric template and the combined password at the same time. The proposed method constructs a fuzzy vault using the feature points extracted from iris. The iris biometric fuzzy vault is then hardened using the password.

Steps in hardening scheme:

1. A combined user and soft biometric password is generated.

2. A random transformation function is derived from the user password.

3. The password transformed function is applied to the Iris template.

4. Fuzzy vault frame work is constructed to secure the transformed templates by using the feature points from the iris.

5. The key derived from the same password is used to encrypt the vault.

Figure 1 depicts the steps involved in the construction of the password hardened multi biometric fuzzy vault.

The organization of the paper is as follows: Chapter II elaborates the background study. Section III explains the proposed generation of cancelable Iris template and securing them using fuzzy vault. Section IV discusses the experimental results and the security analysis. Section V concludes of the proposed work.

## II. RELATED WORK

Karthick Nandakumar et al [5] used the idea of password transformation for fingerprint and generated transformed templates. In his work those transformed templates are protected using fuzzy vault cryptographic construct in which password acts an additional layer of security. Iris based hard fuzzy vault proposed by Srinivasa Reddy [3] followed the same idea of [5] to generate revocable iris templates and secured them using password hardened fuzzy vault. The basic idea of generating cancelable iris is based on the idea derived from the work done by karthick Nandakumar et al [5] and Srinivasa Reddy[3].

Iris based hard fuzzy vault proposed by Srinivasa Reddy [3] applies a sequence of morphological operations to extract minutiae points from the iris texture. This idea is utilized in the proposed method for extracting the minutiae feature point from the Iris. The same idea is used but with combined user and soft biometric password.

Soft biometrics ideas derived from [16, 17, 18, 19, 20] are used for constructing soft biometric passwords.

## III. PROPOSED METHOD

Revocable iris templates generation is carried out in the following three steps. In the first step the iris texture containing the highlighted minutiae feature points is subjected to simple permutation and translation. This results in the original minutiae points being transformed into new points. In the second step the soft biometric password is combined with the user password to get a new 64 bit password. In the third step, the simple transformed iris template is randomly transformed using password. This process enhances the user privacy and facilitates the generation of revocable templates and resists cross matching. This transformation reduces the similarity between the original and transformed template. The transformed templates are further secured using the fuzzy vault construct.

*A. Extraction of Minutiae Feature point from Iris*

The idea proposed by Srinivasa Reddy [3] is utilized to extract the minutiae feature points from the iris texture.

The following operations are applied to the iris images to extract lock/unlock data. Canny edge detection is applied on iris image to deduct iris. Hough transformation is applied first to iris/sclera boundary and then to iris/pupil boundary. Then thresholding is done to isolate eyelashes. Histogram equalization is performed on iris to enhance the contrast. Finally the following sequence of morphological operations is performed on the enhanced iris structure.
(i) closing-by-tophat
(ii) opening

(iii) thresholding

Finally thinning is done to get structures as a collection of pixels. Now the (x, y) coordinates of the nodes and end points of the iris minutiae are extracted. Fig. 2(a) shows the localized iris image, Fig. 2(b) exhibits the iris image with the minutiae patterns and Fig 2(c) shows the permuted and transformed points.

*B. Minutiae Feature Point Transformation*

The Iris texture containing the highlighted minutiae feature points is subjected to simple permutation and translation. This results in the original minutiae points being transformed into new points.

The user password is restricted to the size of 5 characters. The length of the user password is 40 bits. The soft biometric password [16,17,18] is generated by combining height, eye color, and gender. The combination of these three factors results in 24 bit soft biometric password (8 bit each). Therefore the length of the combined password is 64 bits. These 64 bits are divided into 4 blocks of each 16 bits in length.

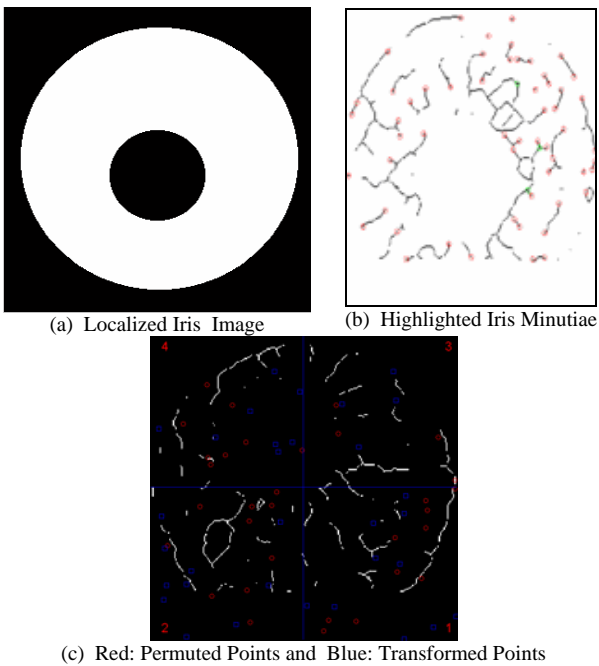

(a) Localized Iris Image   (b) Highlighted Iris Minutiae

(c) Red: Permuted Points and Blue: Transformed Points

Fig 2. Iris Minutiae Extraction and Password Transformation

The feature points highlighted in Iris texture are divided into 4 quadrants. One password block is assigned to each quadrant. Permutation is applied in such a way that the relation position of the minutiae point does not change. Each 16 bit password block is split into two components Tu of 7 bits and Tv of 9 bits in length. Tu and Tv represents the amount of translation in the horizontal and vertical directions, respectively.

The new feature points are obtained by the following transformation.

$$X'_u = (X_u + T_u) \bmod (2^7)$$
$$Y'_v = (Y_v + T_v) \bmod (2^9)$$

where $X_u$ and $X'_u$ is the horizontal distance before and after transformation respectively. Similarly $Y_v$ and $Y'_v$ is the vertical distance before and after transformation respectively.

*C. Fuzzy vault Encoding*

The transformed features are encoded in the fuzzy vault. Password acts as an extra layer of security to the vault. It resists an imposter from modifying the vault. Secret message is generated as a 128 bit random stream. This secret message is transformed with the password. The 16 bit CRC is appended to transformed key S to get 144 bit SC. The primitive polynomial considered for CRC generation is

$$g_{crc}(a) = a^{16} + a^{15} + a^2 + 1$$

The minutiae points whose Euclidian distance is less than D are removed. 16 bit lock/unlock unit 'u' is obtained by concatenating x and y (each 8 bits) coordinates. The 'u' values are sorted and first N of them are selected. The Secret (SC) is divided into 9 non overlapping segments of 16 bits each. Each segment is converted to its decimal equivalent to account for the polynomial coefficients ($C_8$, $C_7$ …$C_0$). All operations takes place in Galois Field GF($2^{16}$).

The projection of 'u' on polynomial 'p' is found. Now the Genuine points set G is ($u_i$, $P(u_i)$). Random chaff points are generated which are 10 times in number that of the genuine points. Both the genuine and chaff point sets are combined to construct the vault. The vault is List scrambled.

*D. Fuzzy vault Decoding*

In the authentication phase, the encrypted vault and bifurcation feature point are decrypted by the combined password. Password based transformation is applied to the query feature points and the vault is unlocked.

From the query templates of the iris, unlocking points (N in number) are extracted. The unlocking set is found as in encoding. This set is compared with the vault to separate the genuine point set for polynomial reconstruction. From this set, all combinations are tried to decode the polynomial. Lagrangian interpolation is used for polynomial reconstruction. For a specific combination of feature points the polynomial gets decoded.

In order to decode the polynomial of degree 8, a minimum of at least 9 points are required. If the combination set contains less then 9 points, polynomial cannot be reconstructed. Now the coefficients and CRC are appended to arrive at SC*. Then SC* is divided by the CRC primitive polynomial.

If the remainder is zero, query image does not match template image and the secret data cannot be extracted. If the remainder is not zero, query image matches with the template image and the correct secret data can be extracted. In this case

SC* is divided into two parts as the 128 bit secret data and 16 bit CRC code.

*E. Parameters used in implementation*

The parameters used in this implementation are shown in Table 1. Chaff points hide the genuine points from the attacker. More chaff points makes the attacker to take much time to compromise the vault but consumes additional computation time. The chaff points added are 10 times in number that of the genuine points.

TABLE I PARAMETERS OF THE IRIS VAULT

| Parameter | Number |
|---|---|
| No. of. Genuine points(r) | 20 |
| No. of Chaff points(c) | 200 |
| Total no. of points (t = r + c) | 220 |

IV. EXPERIMENTAL RESULTS AND ANALYSIS

The Iris template is transformed for three different user passwords to check for revocability. The Table III shows the sample bifurcation points from four quadrants after transformation using three different user passwords and soft biometric passwords generated as shown in table 2 and table 3.

Consider a 5 character user password 'FUZZY', whose ASCII value is given by or 40 bits. Soft biometric password component is 155BM(24 bits). Soft biometric password and User Password are combined to form the transformation password as '155BMFUZZY'(64 bits) whose ASCII values are (155, 66, 77 ,70, 85, 90, 90, 89,) . These 64 bits are divided into four blocks of 16 bits each. Each 16 bit is divided into 7 bits and 9 bits for transformation in horizontal and vertical direction.

The feature point transformation is done with other two user passwords and soft biometric password combinations namely '170GFTOKEN' and '146AM VAULT' whose ASCII codes are (170, 71, 70, 84, 79, 75, 69, 78,) and (146, 65, 77, 86, 65, 85, 76, 84,) respectively. For the same original template different transformed templates are obtained when password is changed. Fig 3(a), Fig 3(b) and Fig 3(c) shows the transformed feature points for three different passwords. The feature points of iris before and after password transformation is shown in table 4. This property of password transformation facilitates revocability. Different password can be utilized for generating different Iris templates.

TABLE II : EYE COLOR AND CHARACTER CODE REPRESENTATION

| Eye Color | Character Code Used |
|---|---|
| Amber | A |
| Blue | E |
| Brown | B |
| Gray | G |
| Green | N |
| Hazel | H |

TABLE III: SHOWING THE STRUCTURE OF SAMPLE PASSWORDS

| User password (5 character) (40 Bits) | Soft Biometric Password (24 bits) | | | Combined Password (64 bits) |
|---|---|---|---|---|
| | Height (0 – 255) (8 bit) | Iris color (1character) (8 bit) | Gender (M/F) (1character) (8 bit) | |
| FUZZY | 155 | B | M | 155BM FUZZY |
| TOKEN | 170 | G | F | 170GF TOKEN |
| VAULT | 146 | A | M | 146AM VAULT |

In the proposed method the security of the fuzzy vault is measured by min-entropy which is expressed in terms of security bits. According to Nanda Kumar [7] the min-entropy of the feature template MT given the vault V can be calculated as

$$H_\infty(M^T | V) = -\log_2 \frac{\binom{r}{n+1}}{\binom{r+c}{n+1}} \quad \ldots\ldots\ldots(1)$$

Where
r = number of genuine points in the vault
c = number of chaff points in the vault
t = the total number of points in the vault (r + c)
n = degree of the polynomial

The security of the iris, retina vault is tabulated in Table. V. In order to decode a polynomial of degree n , (n+1) points are required. The security of the fuzzy vault can be increased by increasing the degree of the vault. Polynomial with lesser degree can be easily reconstructed by the attacker. Polynomial with higher degree increases security and requires lot of computational effort. This makes more memory consumption and makes the system slow. However they are hard to reconstruct.In the case of the vault with polynomial degree n, if the adversary uses brute force attack, the attacker has to try total of (t, n+ 1) combinations of n+ 1 element each. Only (r, n+1) combinations are required to decode the vault. Hence, for an attacker to decode the vault it takes C(t, n+1)/C(r, n+1) evaluations. The guessing entropy for an 8 ASCII character password falls in the range of 18 – 30 bits. Therefore, this entropy is added with the vault entropy. The security analysis of the combined password hardened iris fuzzy vault is shown in Table 5.

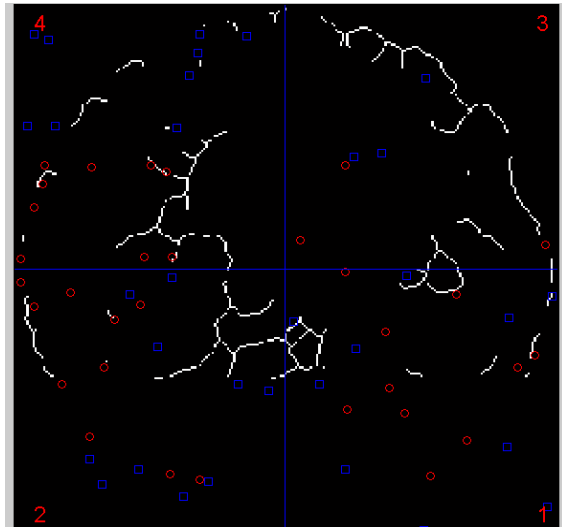

**Fig 3. (a) PASSWORD : VAULT146AM**

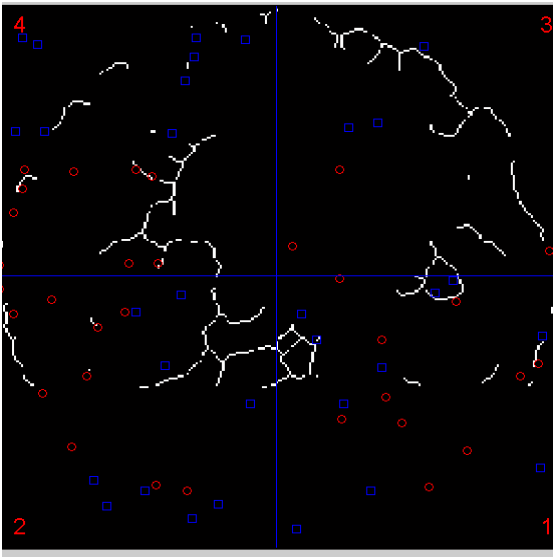

**Fig 3. (b) PASSWORD: FUZZY155BM**

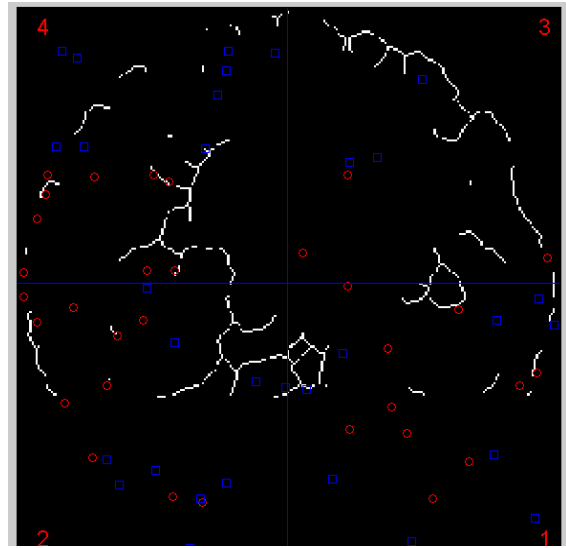

**Fig 3. (c) PASSWORD : TOKEN170GF**

**Fig 3. Transformed Retinal Features for Three Different Soft Biometric Passwords**

TABLE IV   IRIS MINUTIAE FEATURE POINTS BEFORE AND AFTER TRANSFORMATION

| Quadrant and Password | Feature points before transformation | | Transformation code from password | | Feature point after transformation | |
|---|---|---|---|---|---|---|
| | Horizontal Distance (Xu) | Vertical Distance (Yv) | Tu | Tv | Horizontal Distance (Xu) | Vertical Distance (Yv) |
| I '155BM FUZZY' '170GF TOKEN' '146AM VAULT' | 4 | 123 | 77 85 73 | 322 71 65 | 81 89 77 | 61 66 60 |
| II '155BM FUZZY' '170GF TOKEN' '146AM VAULT' | 135 | 114 | 38 35 38 | 326 84 342 | 173 170 173 | 56 70 72 |
| III '155BM FUZZY' '170GF TOKEN' '146AM VAULT' | 4 | 134 | 42 39 32 | 346 331 341 | 46 43 36 | 224 209 219 |
| IV '155BM FUZZY' '170GF TOKEN' '146AM VAULT' | 156 | 129 | 90 69 76 | 89 78 84 | 246 225 232 | 218 207 213 |

TABLE V   SECURITY ANALYSIS OF COMBINED PASSWORD HARDENED FUZZY VAULT

| Vault Type | Degree of polynomia | Min-entropy of the vault (in security bits | Total no: of combinations | Combinations required | No: of Evaluations | Min-entropy + guessing entropy of the password (in security bit) |
|---|---|---|---|---|---|---|
| Iris | 8 | 34 | $2.8187 \times 10^{15}$ | 167960 | $1.6782 \times 10^{10}$ | 52 to 64 |

## V. CONCLUSION

Biometric templates are vulnerable to a variety of attacks. The only disadvantage of biometrics authentication as compared to traditional password based authentication is non revocability. The idea of cancelable biometrics overcomes that disadvantage. Iris has certain advantage as compared to fingerprint. Soft biometrics is ancillary information about a person, when combined with user password gives better results. It is very difficult for an attacker to gain the biometric features, soft biometric components and user password at the same time. The security of these cancelable templates will be the guessing entropy of the 8 character ASCII password which comes to be 18 to 30 bits in strength. The user password can be changed for generating revocable biometric templates. The revocable biometric templates are further secured by applying the ideas of fuzzy vault. Due to this the security of the iris templates increases to 52 to 64 bits.

ACKNOWLEDGEMENT

A public version of the CUHK Iris Database is available from http://www2.acae.cuhk.edu.hk.

*(IJCSIS) International Journal of Computer Science and Information Security,*
*Vol. 7, No. 2, 2010*

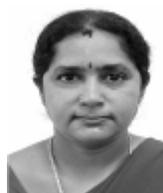

**Dr. Padmavathi Ganapathi** is the Professor and Head of the Department of Computer Science, Avinashilingam University for Women, Coimbatore. She has 21 years of teaching experience and one year Industrial experience. Her areas of interest include Network security and Cryptography and real time communication. She has more than 80 publications at national and International level. She is a life member of many professional organizations like CSI, ISTE, AACE, WSEAS, ISCA, and UWA. She is currently the Principal Investigator of 5 major projects under UGC and DRDO

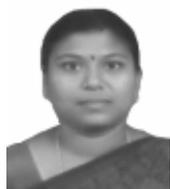

**V. S. Meenakshi** received her B.Sc (Physics) from Madurai Kamaraj University and MCA from Thiagarajar College of Engineering, Madurai in 1990 and 1993 respectively. And, she received her M.Phil degree in Computer Science from Manonmaniam Sundaranar University, Tirunelveli in 2003. She is pursuing her PhD at Avinashilingam University for Women. She is currently working as an Associate Professor in the Department of Computer Applications, SNR Sons College, Coimbatore. She has 16 years of teaching experience. She has presented nearly 15 papers in various national and international conferences. Her research interests are Biometrics, Biometric Template Security and Network Security.